\documentclass[conference]{IEEEtran}
\IEEEoverridecommandlockouts

\usepackage[font=footnotesize,skip=5pt,tableposition=bottom]{caption}
\usepackage{cite}
\usepackage{dblfloatfix}
\usepackage{enumitem}
\usepackage{graphicx}
\usepackage{textcomp}
\usepackage[hyphens]{url}
\usepackage{xcolor}
\usepackage{subcaption}
\usepackage{amsmath}
\usepackage{import}

\def\BibTeX{{\rm B\kern-.05em{\sc i\kern-.025em b}\kern-.08em
		T\kern-.1667em\lower.7ex\hbox{E}\kern-.125emX}}

\renewcommand{\baselinestretch}{0.949}

\newcommand{\opennic}{OpenNIC}
\newcommand{\vscc}{vscc}
\newcommand{\mvcc}{mvcc}
\newcommand{\bmac}{BMac}
\newcommand{\protocolprocessor}{protocol\_processor}

\newcommand{\blkprocessor}{block\_processor}

\newcommand{\txvscc}{tx\_vscc}
\newcommand{\txvalidators}{tx\_validators}

\newcommand{\swpeer}{sw}
\newcommand{\hwpeer}{hw}

\title{Improving Energy Efficiency of Permissioned Blockchains Using FPGAs} 
\author{
	\IEEEauthorblockN{
		Nathania Santoso \enspace
		Haris Javaid
	}
	\IEEEauthorblockA{
		AMD, Singapore
	}
	\IEEEauthorblockA{\{nathania.santoso, haris.javaid\}@amd.com}
}

\begin{document}
\maketitle

\pagestyle{plain}
	
\begin{abstract}
Permissioned blockchains like Hyperledger Fabric have become quite popular for implementation of enterprise applications. Recent research has mainly focused on improving performance of permissioned blockchains without any consideration of their power/energy consumption. In this paper, we conduct a comprehensive empirical study to understand energy efficiency (throughput/energy) of validator peer in Hyperledger Fabric (a major bottleneck node). We pick a number of optimizations for validator peer from literature (allocated CPUs, software block cache and FPGA based accelerator). First, we propose a methodology to measure power/energy consumption of the two resulting compute platforms (CPU-only and CPU+FPGA). Then, we use our methodology to evaluate energy efficiency of a diverse set of validator peer configurations, and present many useful insights. With careful selection of software optimizations and FPGA accelerator configuration, we improved energy efficiency of validator peer by 10$\times$ compared to vanilla validator peer (i.e., energy-aware provisioning of validator peer can deliver 10$\times$ more throughput while consuming the same amount of energy). In absolute terms, this means 23,000 tx/s with power consumption of 118W from a validator peer using software block cache running on a 4-core server with AMD/Xilinx Alveo U250 FPGA card.

\end{abstract}


\section{Introduction} \label{introduction}
Blockchain technology is on the rise due to its capability of executing transactions (which contain business logic in the form of smart contracts) and storing them in a decentralized manner (same ledger distributed among multiple nodes). The distributed ledger contains blocks where each block has a hash value of itself and the previous block, assuring immutability of the ledger data. Hence, a network of nodes that implement a blockchain essentially provides an implementation of a distributed ledger for applications that will be developed and deployed on top of that blockchain. There are two types of blockchains: public (permissionless) and private (permissioned) blockchains. In a public blockchain, nodes do not require identity authorization to participate in the network. On the other hand, the identity of a node must be authenticated cryptographically to execute transactions in a permissioned blockchain. Bitcoin and Ethereum are examples of public blockchains, while Hyperledger Fabric~\cite{HyperledgerFabric} is an example of permissioned blockchain.

Hyperledger Fabric is one of the most popular permissioned blockchains for enterprise applications~\cite{Castillo2021}. Most of the recent research on Fabric has only focused on improving its throughput (transactions per second, shortened as tx/s), overlooking its power and energy consumption. Sedlmeir et al.~\cite{Sedlmeir2020BlockchainEnergy} estimated that although a Fabric network consumed 8 orders of magnitude less energy than a public blockchain, it still consumed 10$\times$ more energy than a centralized system. Furthermore, Fabric nodes are often deployed in datacenters where energy consumption has become an issue due to environmental and climate change concerns. Therefore, there is a need to explore energy efficiency of Fabric nodes. Previous works~\cite{Gorenflo2019FastFabric,Thakkar2021ScalingBlockchains,Javaid2022BlockchainMachine} have shown that validator peer node (which runs the validation of blocks and transactions before committing them to the ledger) is one of the major bottlenecks. It is a computationally intensive operation, and thus utilizes significant resources which will result in high power and energy consumption in the Fabric network.

In this paper, we propose a power/energy measurement methodology to conduct a comprehensive evaluation of Hyperledger Fabric validator peer in terms of its power/energy consumption and throughput. Typically, Fabric peer is deployed on a multicore server (i.e., CPU based system). Recently, Fabric peer has also been shown to run on a multi-core server with hardware accelerator on an FPGA card (i.e., CPU+FPGA based system) for high throughput~\cite{Javaid2022BlockchainMachine}.
In this context, \textbf{this paper has the following contributions:}
\begin{itemize}
	\item We propose a power/energy measurement methodology for a CPU based system, where our approach handles varying number of vCPUs allocated from the physical cores of a multi-core server. We use this setup to evaluate energy efficiency of vanilla validator peer and a prominent software optimization (block cache from~\cite{Gorenflo2019FastFabric}).
	\item We integrate a power measurement module inside the hardware accelerator from~\cite{Javaid2022BlockchainMachine} for power/energy measurement of a CPU+FPGA based system. We use this setup to evaluate energy efficiency of hardware accelerated validator peer with and without the software block cache.
	\item We finally present a comprehensive study of the interactions between power, energy and throughput of validator peer, and deduce many insights from comparison of CPU-only and CPU+FPGA systems for energy-aware provisioning of validator peers in a Fabric network.
\end{itemize}

Our experiments with Hyperledger Fabric v2.2 LTS running on a multi-core server with Alveo U250 card show that FPGA based hardware accelerator combined with software block cache can deliver 10$\times$ more throughput than a CPU-only system with same energy consumption (153 vs. 15 tx/s/J).

\section{Background and Preliminaries}\label{background_and_preliminaries}
\begin{figure}[t!]
	\centerline{\includegraphics[scale=0.95]{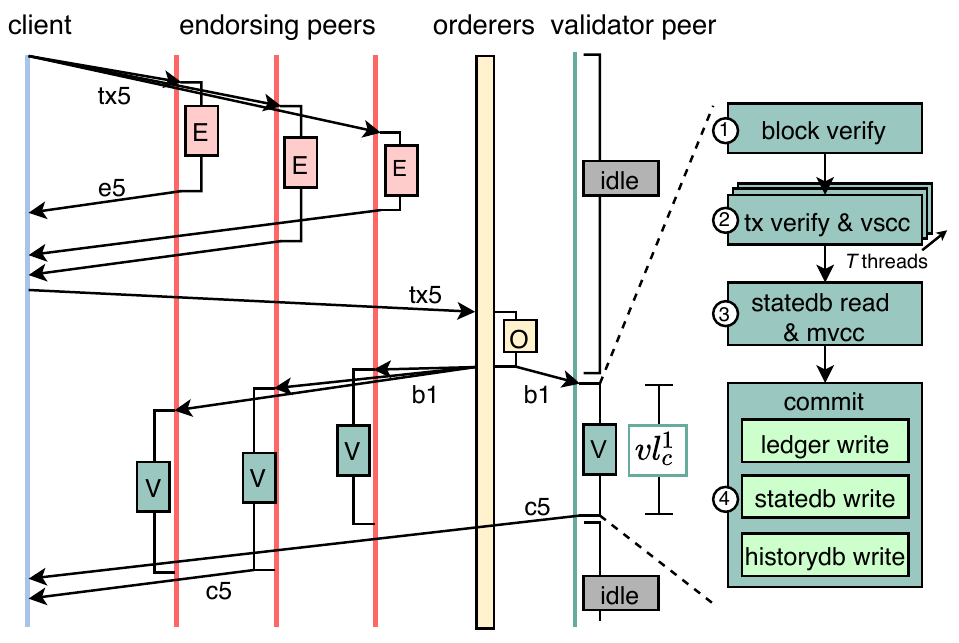}}
	\caption{Transaction flow in Hyperledger Fabric. txN = transaction N, eN = endorsement for txN, bN = block N, and cN = confirmation for txN. E = endorsement, O = ordering, and V = validation phases.}
	\label{fabric_txflow}
	\vspace{-2ex}
\end{figure}

\subsection{Hyperledger Fabric}\label{hyperledger_fabric}
Hyperledger Fabric is an open-source and enterprise-grade implementation of a permissioned blockchain. It is not associated with any cryptocurrency, and has applications in many diverse domains such as supply chain, banking/finance, healthcare, etc. Figure~\ref{fabric_txflow} depicts how a transaction flows through various nodes of a Fabric network (e.g., endorsing peers, validator peers, etc. shown on the top).  

A client invokes a transaction by sending it to endorsing peers. Each endorsing peer will simulate the transaction by computing its input and output (read and write sets) against the locally stored state database\footnote{The state database contains the current snapshot of the ledger. For example, for a banking application, it would contain the current value of each account.}. Afterwards, the endorsing peer will send the simulation results along with its digital signature back to the client. Once the client has enough endorsements, it will send the transaction along with the endorsements to the orderer for inclusion into a block. The orderer will create a new block and broadcast it to all the peers for validation and subsequent commit to the ledger. Note that each transaction contains the digital signature of the client that invoked it, while each block contains the digital signature of the orderer that created it. In Fabric, all digital signatures are based on Elliptic Curve Digital Signature Algorithm (ECDSA) scheme.

The validation and commit phase itself consists of several operations as shown on the right hand side of Figure~\ref{fabric_txflow}. When a validator peer receives a block, it will first check the syntax of the block and verify the orderer's signature on the block (step 1). Then, the validator peer will check the syntax of each transaction in the block and verify the client's signature on each transaction (tx verify in step 2). Afterwards, for each transaction in the block, the signatures of endorsing peers from the transaction's endorsements will be verified and evaluated against an endorsement policy\footnote{An endorsement policy is associated with a smart contract/chaincode, and governs the business logic for approval of the transaction; e.g., policy of a money transfer chaincode between two banks may require valid endorsements from each bank, i.e., endorsements from Bank1 AND Bank2.} (validation system chaincode, shortened as tx \vscc{} in step 2). If the endorsement policy is satisfied, then the transaction is marked as valid.


In step 3, multi-version concurrency control (shortened as \mvcc{}) checks are applied to mark a transaction as valid/invalid. Once the entire block has been validated in steps 1--3, in step 4, the validator peer will commit the block to its ledger, and update its state database (by applying write sets of valid transactions) and history database (for book keeping). Note that validator peers only validate and commit incoming blocks, while endorsing peers do the same in addition to endorsing incoming transactions.

The vanilla Fabric validator peer incorporates many software optimizations for improved throughput. For example, it uses multiple threads to verify and validate multiple transactions of a block in parallel (taking advantage of multi-core servers), which is depicted as \textit{T threads} in Figure~\ref{fabric_txflow}. Another software optimization is to use \textit{block cache} which caches unmarshaled contents of a block for subsequent accesses, and has been reported to improve throughput significantly (2.33$\times$ in~\cite{Gorenflo2019FastFabric} and 1.67$\times$ in~\cite{Thakkar2021ScalingBlockchains}). However, block cache is not yet part of the official Fabric codebase, so we implemented it ourselves in Fabric v2.2 LTS to evaluate its energy efficiency.

\subsection{Blockchain Machine}\label{blockchain_machine}
Javaid et al.~\cite{Javaid2022BlockchainMachine} proposed a hardware accelerator called \textit{Blockchain Machine (\bmac{})} for improving validator peer's throughput,  instead of relying on software optimizations. The BMac peer is designed for a multi-core server with a network-attached FPGA card (connected to the CPU via PCIe bus), where the CPU runs modified Fabric software while FPGA card is programmed with the hardware accelerator. Since this is the only hardware accelerator proposed for Hyperledger Fabric so far, we use it as the CPU+FPGA system in this paper.

\begin{figure}[t!]
	\centerline{\includegraphics[scale=0.75]{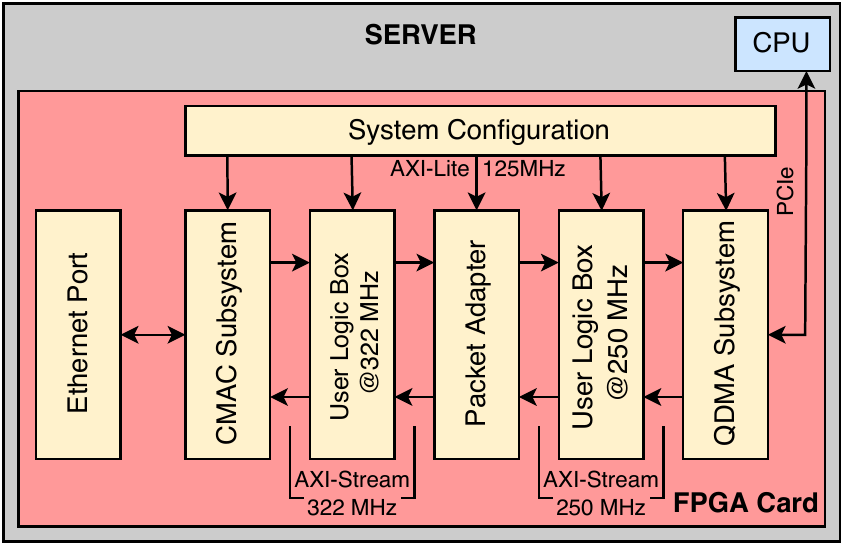}}
	\caption{Overview of OpenNIC shell. The Blockchain Machine hardware is part of User Logic Box @ 250MHz.}
	\label{opennicshell_arch}
	\vspace{-4ex}
\end{figure}

Figure~\ref{opennicshell_arch} provides a simplified overview of the open-sourced \opennic{} shell~\cite{Xilinx2020OpenNIC}, which is the basis for \bmac{} hardware. The OpenNIC shell is an FPGA based NIC shell for AMD/Xilinx FPGA cards and provides network connectivity through CMAC/Ethernet port interfacing and CPU connectivity through QDMA/PCIe interfacing. Consequently, a user-defined accelerator can be implemented inside the user logic box where it can access incoming data from the network through CMAC/Ethernet port while the CPU can access the output of hardware accelerator through PCIe/QDMA.

\begin{figure}[t!]
    \centerline{\includegraphics[scale=0.93]{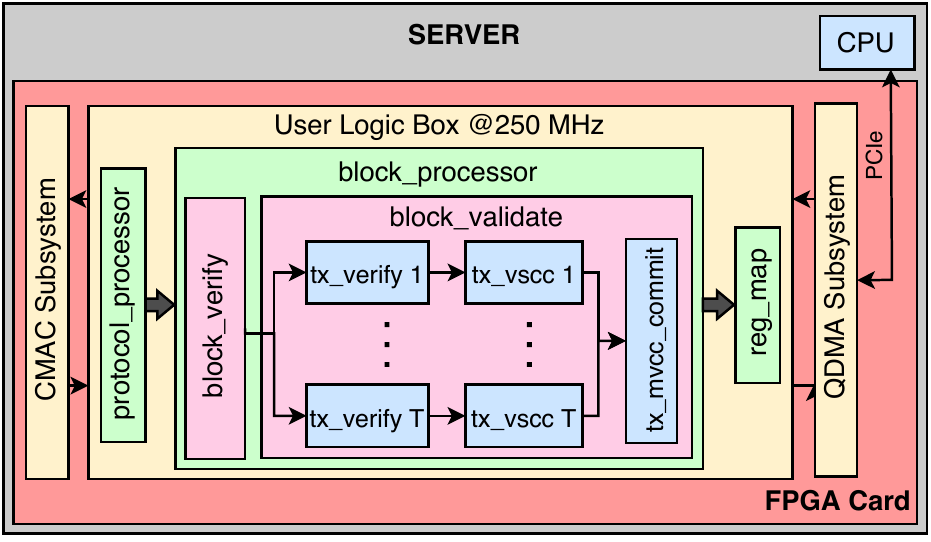}}
	\caption{Overview of Blockchain Machine hardware. Only the relevant modules of OpenNIC shell are shown here.}
	\label{bmac_arch}
	\vspace{-4ex}
\end{figure}

The \bmac{} hardware is implemented as user logic box @ 250MHz, as shown in Figure~\ref{bmac_arch}. The blocks are received in FPGA card through the CMAC interface. The first module, \protocolprocessor{}, processes the incoming Ethernet packets and extracts relevant data, such as block id, transaction ids, ECDSA signatures, etc. The second module, \blkprocessor{}, uses this data to validate the block and its transactions, commits all valid transactions, and then writes the validation results in a register map (for CPU access through QDMA).

\begin{figure}[b!]
    \vspace{-4ex}
	\centerline{\includegraphics[scale=0.95]{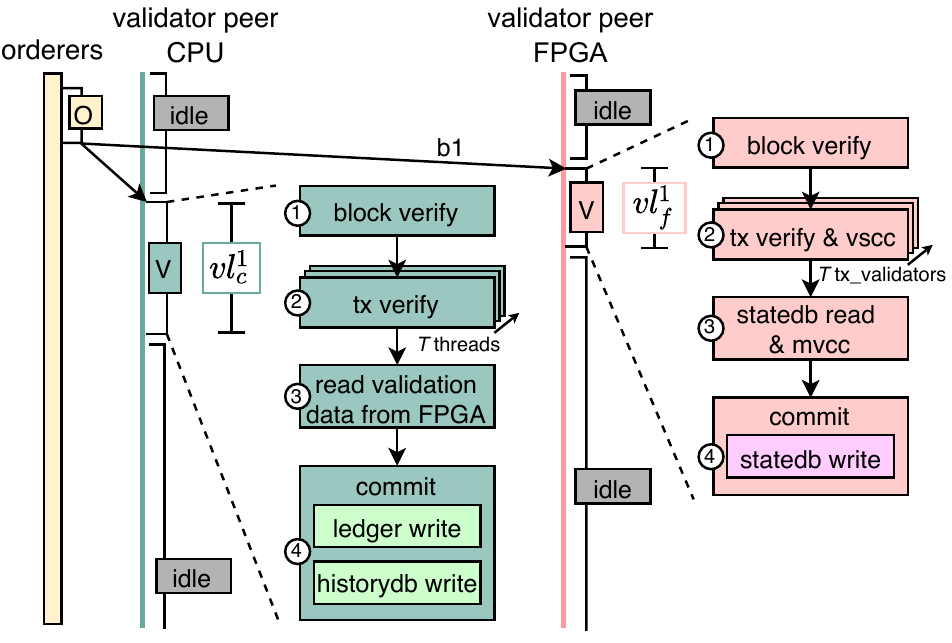}}
	\caption{Validator peer with Blockchain Machine hardware.}
	\label{fabric_txflow_wbmac}
\end{figure}

Figure~\ref{fabric_txflow_wbmac} shows how blocks and transactions are processed in \bmac{} peer. The \bmac{} hardware validates the block without any involvement of the CPU (right hand side). The same block is also received by the Fabric peer software running on the CPU (left hand side), which still executes some parts of the block and transaction verification that are not suitable for hardware accelerator. After that, the software skips validation operations and just reads validation results of the block from hardware, combines them with the original block, and then commits the updated block to ledger just like any other validator peer in the Fabric network. In other words, the validation phase is offloaded to the network-attached hardware accelerator on FPGA.

Note that the \blkprocessor{} in \bmac{} hardware has several pipeline stages and a configurable number of parallel validators, which is shown as \textit{T \txvalidators{}} in Figs.~\ref{bmac_arch} \&~\ref{fabric_txflow_wbmac}. Consequently, the \blkprocessor{} processes multiple transactions in a parallel-pipelined fashion.

\section{Proposed Power Measurement Methodology}\label{power_measurement_methodology}
\subsection{Motivation and Challenges}\label{motivation_and_challenges}
The Fabric validator peer with software optimizations has been shown to achieve a throughput of 14,000 tx/s~\cite{Gorenflo2020FastFabricImplementation}, while the hardware accelerator has been shown to achieve a throughput of up to 69,000 tx/s~\cite{Javaid2022BlockchainMachine}. Although these throughput numbers are impressive, all the previous works overlook energy efficiency of the validator peer. Therefore, we aim to conduct a comprehensive study of not only the validator peer's throughput but also its power/energy consumption. We list several challenges that exist when measuring power/energy in multi-core servers with an FPGA card, and then present our methodology to handle those challenges.

\textbf{Challenge 1: }
The ideal method for measuring power of a server is to use a power meter to measure what is called the wall power, and then compute energy consumption. However, multi-core servers are typically housed in datacenter-like environments where physical access is limited. Therefore, we use hardware counters available in modern processors to measure energy consumption, just like many previous works do~\cite{Khan2018RAPL}. The issue with this approach is that such hardware counters are available at various granularities across different types of processor architectures. In some architectures, they are available for each core, while in other architectures, they are only available for each socket containing multiple cores which makes it impossible to directly get the energy consumption of each core. The availability of per-core energy consumption is pertinent because typical multi-core servers contain tens of physical cores where a subset of these cores is provisioned for a particular application as virtual CPUs (vCPUs)~\cite{Krzywda2018DatacenterPowerPerformance}. For example, a validator peer may be allocated only 8 vCPUs when running on a server with 32 physical cores. The challenge here is to measure energy consumption of a validator peer provisioned with varying number of vCPUs when per-core energy consumption is not available from hardware counters.

We propose a method in Section~\ref{cpu_idle_power} where we stress different number of vCPUs in a server to empirically deduce power consumption of an idle core. Then, we use this idle core power to adjust the overall energy consumption of a validator peer. Note that our method can be skipped and replaced with per-core energy consumption values when they are directly available from hardware counters.

\textbf{Challenge 2: }
The validator peer receives and processes a block, and then waits for the next block as shown in Figs.~\ref{fabric_txflow} \&~\ref{fabric_txflow_wbmac}. The idle period between blocks depends on how fast an orderer can form a block, which further depends on the aggregate rate at which clients generate/send transactions. More clients sending transactions at higher rates will result in overall higher transaction send rate, which will translate to shorter idle periods between blocks.

Naively measuring energy consumption when idle periods are much longer than block validation time means that the energy consumption would be dominated by the idle power of cores which will mislead the analysis. Furthermore, any energy savings from speeding up the validation phase will not be apparent. Therefore, in an ideal setup, validator peers should be saturated, i.e., aggregate transaction send rate is high enough to result in minimal idle periods. However, saturating validator peers in a research/experimental environment may not be possible due to lack of high-end servers that can run tens of clients, meaning that there is not enough computational power to generate the required transaction workload. Hence, the challenge here is to automatically detect and exclude idle periods from the execution of a validator peer irrespective of the clients' aggregated transaction send rate and orderer's block creation rate.

We propose a method in Section~\ref{cpu_idle_period} which collects execution log of a validator peer and energy consumption log of the server while validator peer is running. We overlay these logs to detect and exclude idle periods for computation of total energy consumption, as if the validator peer was saturated.

\textbf{Challenge 3: }
For the validator peer with hardware accelerator, we must measure energy consumption of both the CPU and FPGA because lightweight operations in Fabric software run on the CPU while computationally intensive operations run on the FPGA. For CPU energy consumption, we use the method proposed in Section~\ref{cpu_power_measurement}. However, for FPGA power/energy consumption, the \opennic{} shell does not provide any mechanism. 

We use AMD/Xilinx Card Management Solution (CMS) IP to measure FPGA card power, and integrate it into \opennic{} shell as an additional subsystem. We describe the architecture of CMS IP, its integration, and its access from CPU in Section~\ref{fpga_power_measurement}. Finally, we propose our complete methodology in Section~\ref{overall_measurement} which takes care of intricacies between CPU and FPGA validation phases to compute validator peer's overall throughput and energy/power consumption for both CPU-only and CPU+FPGA systems.

\subsection{CPU Power Measurement}\label{cpu_power_measurement}
\subsubsection{Intel Hardware Counters}\label{intel_cpu_power}
Modern Intel processors provide the Running Average Power Limit (RAPL)~\cite{Khan2018RAPL} interface for fine-grained measurement of energy consumption. The RAPL interface supports multiple power domains, e.g. package, PP0 and DRAM domains report energy consumption of an entire socket, only the cores in a socket, and only the connected DRAMs, respectively. The RAPL interface uses Model Specific Registers (MSRs) which are basically hardware counters for accumulated energy consumption of different modules, and hence not all domains are available on all processor architectures~\cite{Intel2019SoftwareDeveloperManualVol4}. For example, Sandy Bridge supports PP0 domain while Haswell-EP does not. These MSRs are updated approximately every 1ms \cite{Khan2018RAPL}.

One can read energy consumption values directly from MSRs, or using \textit{sysfs} interface or \textit{perf} command on Linux. We use perf command in this paper because it abstracts away differences in processor architectures and provides a simple command-line interface instead. We conducted many experiments measuring energy consumption directly from MSRs and perf command, and found the difference to be always within 1.5\%. For all our measurements, we include both the CPU and DRAM energy consumption.

\subsubsection{CPU Idle Power}\label{cpu_idle_power}
As explained in Section~\ref{motivation_and_challenges}, if per-core energy consumption is not available through a RAPL domain, then we need a mechanism to measure idle power of a core which can later be used to adjust socket energy consumption (RAPL package domain) for a more realistic measurement (Sec.~\ref{overall_measurement}).

Our approach to measure idle core power is as follows. We measure socket energy consumption when there is no workload running on the server (all cores are idle). Then, we generate workload for only 1 core using \textit{stress} command on Linux and measure socket energy consumption again. We repeat this process by generating workload for 1 more core every time until all the physical cores are used. Since stress command fully utilizes a core (100\% CPU utilization), in each step, we increase active cores by 1 while decreasing idle cores by 1. Hence, in this controlled setup, we expect a linear relationship $P_{s} = C_{a} \times P_{c}^{a} + C_{i} \times P_{c}^{i}$ where $C_{a}$ and $C_{i}$ are the number of active and idle cores respectively, while $P_{s}$, $P^{a}_{c}$ and $P^{i}_{c}$ are the power consumption of the entire socket, an active core and an idle core respectively. In our approach, $P_{s}$ is measured as socket energy consumption divided by the measurement interval while $C_{a}$ and $C_{i}$ are known. Hence, we use liner regression to fit the measured data to deduce the values of $P_{c}^{a}$ and $P_{c}^{i}$. We ran the stress commands natively on the server as well as within VMs which are provisioned with 1 vCPU, 2 vCPUs and so on. All our measurements are for an interval of 30 seconds. Based on this setup, the estimated idle core power $P_{c}^{i}$ = 1.44W in our servers. Note that the number of vCPUs we provision in our setup is always less than or equal to the total number of physical cores, thus we avoid the use of hyperthreading which is known to make power/energy measurements less accurate~\cite{Yan2014Happy}.

\subsubsection{CPU Idle Periods}\label{cpu_idle_period}
As explained in Section~\ref{motivation_and_challenges}, we need a mechanism to detect idle periods in a validator peer so they can be excluded when energy consumption is computed (in order to imitate a saturated validator peer regardless of the clients' aggregated transaction send rate or orderer's block creation rate).

Our approach is as follows. We collect execution log of a validator peer where we have modified the Fabric codebase to log timestamps for various operations, e.g. vscc, mvcc, etc. At the same time, we collect socket energy consumption log with timestamps of the server by running perf command with an interval of 10ms. We chose 10ms because it provided fine-grained granularity to measure energy consumption across all operations of the validation phase. Since both these logs have timestamps, we overlay the execution log on the energy consumption log to create an annotated time-series data, that is, we add markers in the energy consumption log to indicate start and end of each validation operation. Afterwards, we analyze the annotated time-series data to compute total energy consumption of the validator peer while excluding idle periods.

\begin{figure}[t!]
	\centerline{\includegraphics[scale=0.5]{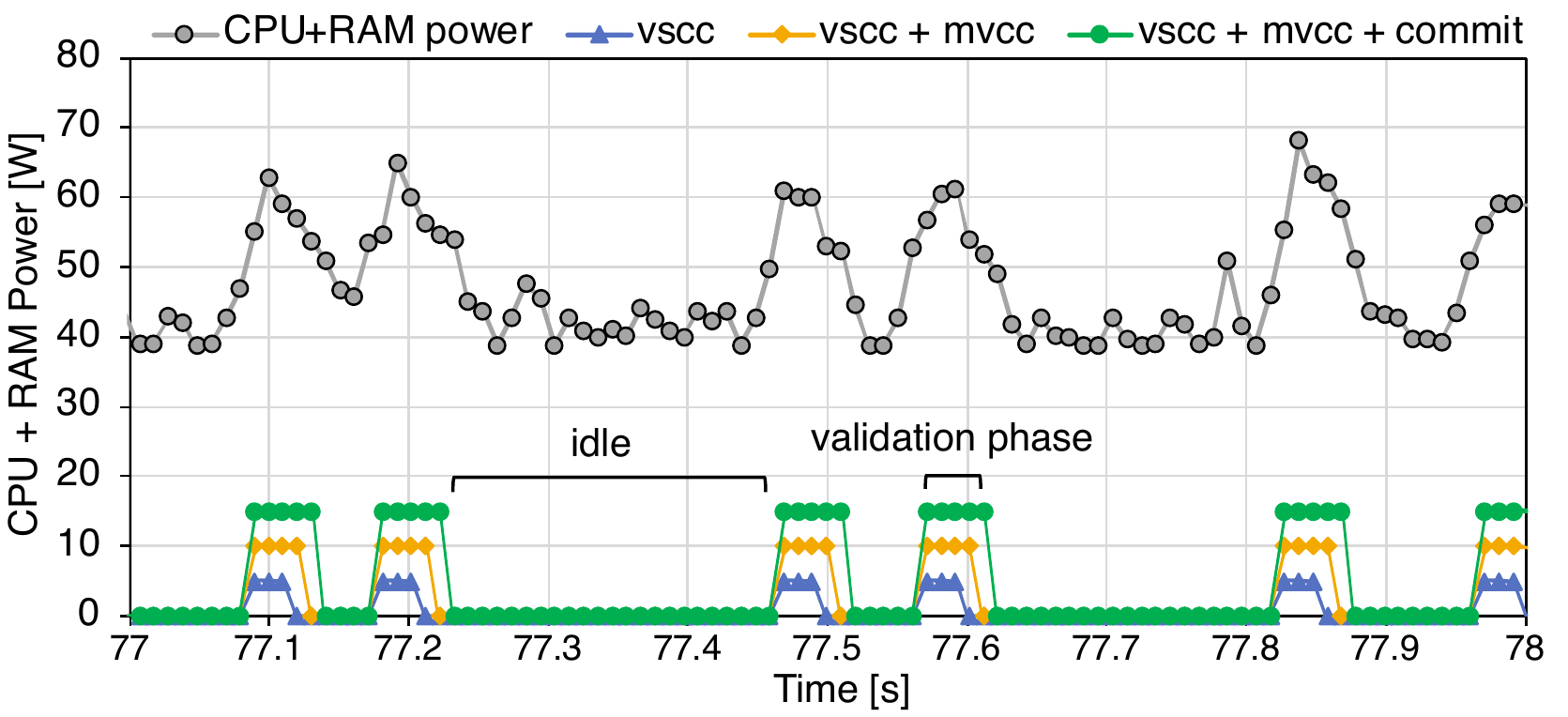}}
	\caption{Annotated time-series power measurements of validator peer.}
	\label{eliminate_idle}
	\vspace{-4ex}
\end{figure}

Figure~\ref{eliminate_idle} shows the annotated time-series power measurements (each energy consumption measurement divided by 10ms) for validator peer with markers added for vscc, mvcc and commit operations. When there are no markers, then there are no validation operations running and the validator peer is waiting for the next block. It is clear that power consumption increases significantly when blocks are being validated, while the power consumption is minimal during idle periods. Our peer runs in a VM (where execution log is captured) which is time-synchronized with the host server (where energy consumption log is captured), hence we use scripts to automatically overlay these logs, detect and exclude idle periods to compute total energy consumption. Note that even when a validator peer is saturated with enough transaction workload, our approach is still applicable since we only analyze idle periods from the execution log.

\subsection{FPGA Power Measurement}\label{fpga_power_measurement}
Modern FPGA cards have on-board voltage and current sensors to enable power measurements. Several power domains are supported (e.g., total card power, only FPGA power, only BRAM power, etc.) but not all of them are available in every FPGA card. AMD/Xilinx provides Card Management Solution (CMS) IP~\cite{Xilinx2022Pg348} that interacts with these sensors to report power consumption of the corresponding domains.

\begin{figure}[t!]
	\centerline{\includegraphics[scale=0.8]{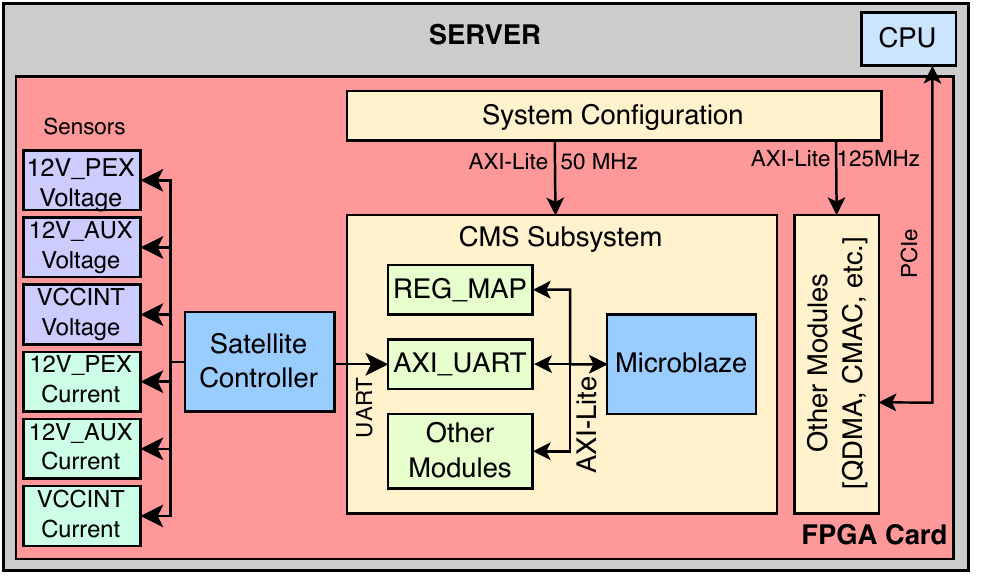}}
	\caption{Integration of CMS IP into \opennic{} shell for power measurements.}
	\label{cms_ip_arch}
	\vspace{-3ex}
\end{figure}

Figure~\ref{cms_ip_arch} shows how we integrate the CMS IP into the \opennic{} shell as an additional subsystem. Internally, the CMS subsystem consists of a Microblaze processor that runs a firmware that interacts with an on-board satellite controller through UART. The satellite controller accesses on-board sensors to read voltage and current values for different power domains. The Microblaze firmware polls the satellite controller approximately every 120ms to read the new values, and writes them into a shared memory (REG\_MAP) for access by host CPU (server). The CMS subsystem is connected to \opennic{} shell's QDMA subsystem through the system configuration module using AXI-Lite interface. As a result, the host CPU can access REG\_MAP of CMS subsystem over PCIe to read power measurements reported by the satellite controller.

We use AMD/Xilinx Alveo U250 FPGA card in our experiments. The U250 card supports either the total card power domain (12V\_PEX and 12V\_AUX voltages/currents) or only FPGA power domain (VCC\_INT voltage/current)~\cite{Xilinx2022Pg348}. We use the total card power as power consumption of the hardware accelerator even when we use only CPU and DRAM energy consumption for CPU based system (instead of total motherboard power). In other words, total card power is an upper-bound on the power consumed by the hardware accelerator.

To compute energy consumption of \bmac{} hardware, we collect FPGA power consumption log by running a script on the host that reads and logs total card power with timestamps. The Fabric peer software running on the CPU reads and logs latency to validate each block from \bmac{} hardware registers in its execution log (in addition to the logging described in Sec.~\ref{cpu_idle_period}). Since both these logs have timestamps, we use scripts to automatically merge them to compute energy consumption of the validation phase on FPGA. Note that the smallest interval for FPGA power measurements is 120ms (due to CMS IP limitation), so to be consistent with 10ms CPU energy measurement interval in Sec.~\ref{cpu_idle_period}, we assume the same power consumption for all the occurrences of validation phase in the 120ms interval.

\subsection{Bringing It All Together: Validator Peer Throughput and Energy Consumption}\label{overall_measurement}
In this section, we describe how we put together all the proposed approaches to compute validator peer's throughput and total energy consumption. We use the following terms (which are all measured in seconds):
\begin{itemize}
    
    \item $vl_{c}^{i}$ and $vl_{f}^{i}$: validation latency of i-th block on CPU and FPGA respectively (Fig.~\ref{fabric_txflow} and Fig.~\ref{fabric_txflow_wbmac} respectively). 
    \item $tt_{c}$ and $tt_{f}$: total time to validate all blocks on CPU and FPGA respectively (computed as sum of $vl_{c}^{i}$ and $vl_{f}^{i}$ respectively for all blocks).
    \item $tt$: total time to validate all blocks in a validator peer, computed as $max(tt_{c}, tt_{f})$ because validation phase executes on both CPU and FPGA in parallel.
\end{itemize}
The throughput of a validator peer is computed as total transactions in all blocks divided by $tt$, which is essentially the rate at which transactions are committed by the peer. The total energy consumption and average power consumption of a validator peer are computed as:
\begin{equation}
\begin{split}
    E =& E_{c} + E_{f} = E_{vp} + E_{ip} - E_{cores}^{u} + E_{f} \nonumber \\
    P =& E / tt 
\end{split}
\end{equation}

where $E_{c}$ and $E_{f}$ denote energy consumption of CPU and FPGA respectively, and the remaining terms are defined as:
\begin{itemize}
    \item $E_{vp}$: Energy consumption of the validation phase on CPU, excluding idle periods. It is computed as described in Sec.~\ref{cpu_idle_period}, and is essentially the energy consumption of CPU and DRAM over the $tt_{c}$ period.
    \item $E_{ip}$: Energy consumption of CPU and DRAM when the CPU is waiting for the FPGA (i.e., $tt_{f} > tt_{c}$ and $tt = tt_{f}$). It is computed as $P_{ip} \times (tt - tt_{c})$ where $P_{ip}$ is computed as the average CPU + DRAM power from all the CPU idle periods (e.g., about 40W in Fig.~\ref{eliminate_idle}). This term ensures that CPU and DRAM energy consumption is included even when CPU is idle and waiting for the next block due to slower validation phase on FPGA.
    \item $E_{cores}^{u}$: Energy consumption of unused physical cores of a socket, that is, cores that are not provisioned as vCPUs for the validator peer. It is computed as $P_{c}^i \times C_{i} \times tt$ where $P_{c}^i$ and $C_{i}$ are computed as described in Sec.~\ref{cpu_idle_power}. As an example, if a socket has 10 physical cores and validator peer is run in a VM with 8 vCPUs (pinned to this particular socket), then the idle power of 2 cores is subtracted to compute a more realistic energy consumption of the validator peer.
    \item $E_{f}$: Energy consumption of the validation phase on FPGA, computed as described in Sec.~\ref{fpga_power_measurement} over the $tt$ period. Note the use of $tt$ period instead of $tt_{f}$ to ensure that FPGA energy consumption is included even when its idle and waiting for the next block due to slower validation phase on CPU.
\end{itemize}

Note that when validator peer is run on a CPU-only system, all the FPGA related terms ($tt_{f}$, $E_{f}$, etc.) are zero, and hence all the above equations degenerate into simpler CPU-only equations. For direct comparison between CPU-only and CPU+FPGA systems, we exclude ledger write latency from $vl_{c}^{i}$ because ledger write operation is always executed on CPU, and can either be executed asynchronously in the background or on a separate storage server~\cite{Gorenflo2019FastFabric}.

Our overall approach is algorithmically described below, and is implemented using fully automated scripts:
\begin{itemize}
    \item Gather validator peer's execution log, socket energy consumption log and fpga power consumption log.
    \item Process the above logs to compute $tt_{c}$, $tt_{f}$ and $tt$.
    \item Process execution and socket energy consumption logs to compute $E_{vp}$ and $P_{ip}$, and thus $E_{ip}$.
    \item Process execution and fpga power consumption logs to compute $E_{f}$.
    \item Compute validator peer's throughput, total energy consumption and average power consumption.
\end{itemize}

\section{Evaluation Setup \& Experimental Results}\label{evaluation_setup}
\subsection{Hyperledger Fabric Network and Application Setup}\label{fabric_network_setup}
We create a Fabric network with two organizations where each organization has 1 endorsing peer and 1 validator peer, and a solo RAFT orderer. These organizations interact with each other through the smallbank smart contract~\cite{HyperledgerCaliperBenchmarks} where AND endorsement policy is configured, meaning that all transactions must be approved by both organizations. This is a typical setup used in many previous works~\cite{Gorenflo2019FastFabric,Thakkar2021ScalingBlockchains,Zhu2020}, and is representative of two banks processing banking transactions. We use Hyperledger Caliper~\cite{HyperledgerCaliper}, the standard blockchain benchmarking tool, with in-house scripts to automatically bring up the Fabric network, generate transaction workload (30,000 random transactions), collect logs, and report throughput and energy/power consumption of the validator peers.

\subsection{Hardware/Software Setup}\label{hw_sw_setup}
We use dual-socket servers where each socket has 10 Intel Xeon 4114 @ 2.2GHz cores. Each peer is run in its own VM which is provisioned with a certain number of vCPUs and 1GB RAM per vCPU, where these vCPUs are pinned to physical cores. The pinning process allows us to measure energy consumption more accurately. For example, we can create two VMs each with 8 vCPUs pinned to physical cores from two separate sockets. As such, these VMs will not interfere with each other and their energy consumptions can be measured by collecting energy consumption of the sockets separately. Likewise, if vCPUs of a VM are pinned to physical cores in both sockets, then we only create a single VM in that server. Both the orderer and Caliper are run in their own VMs with 8 vCPUs. We used Fabric v2.2 with LevelDB for each peer, and the number of vscc threads is the same as vCPUs.

All our experiments use the Fabric network from above where organization 1's validator peer is run on a VM (CPU-only system) and then run again on a VM with FPGA card (CPU+FPGA system). The \bmac{} hardware on FPGA card is configured with varying number of \txvalidators{} where 2 ECDSA engines are used in each \txvscc{} instance for AND endorsement policy~\cite{Javaid2022BlockchainMachine}.

\vspace{-1ex}
\subsection{Evaluation Metrics}\label{evaluation_metrics}
The primary performance and energy efficiency metrics we use are throughput (the rate at which peer commits transactions, measured as tx/s) and throughput/energy (transactions committed per second while consuming 1 Joule of energy, measured as tx/s/J) respectively. In some cases, we also report the average power consumption and highlight why energy consumption is more relevant than just the power consumption. All these metrics are computed as described in Section~\ref{overall_measurement}.

\begin{figure}[t!]
	\centering
	\includegraphics[scale=0.47]{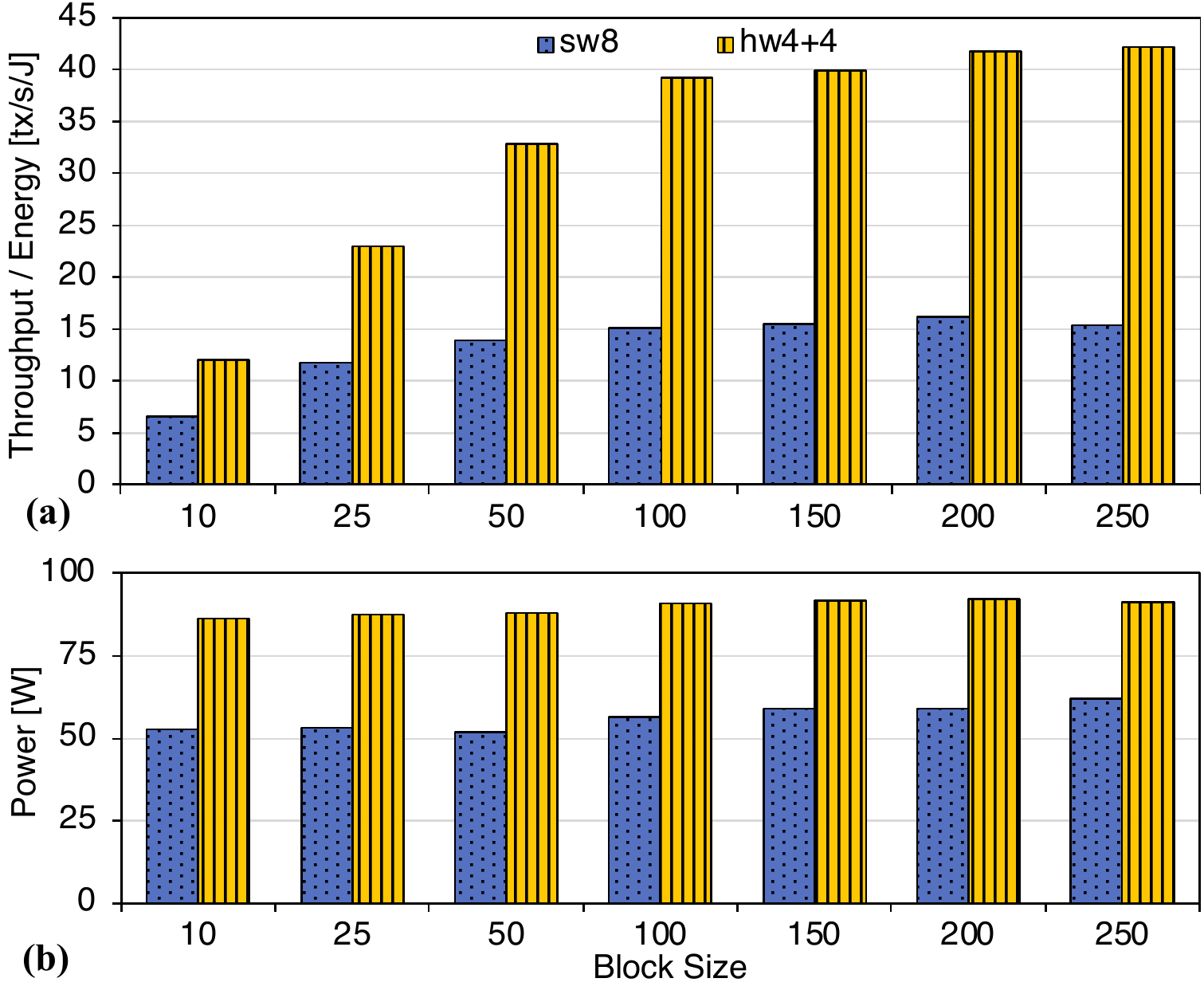}
	\caption{(a) Energy efficiency, and (b) power consumption of a \swpeer{} and \hwpeer{} peer.}
	\label{blocksize_variation}
	\vspace{-4ex}
\end{figure}

\subsection{Validator Peer Energy Efficiency}
We first create two configurations to understand the energy efficiency of validator peer:
\begin{itemize}
    \item \textbf{\swpeer{}\textit{N}}: vanilla validator peer using N vCPUs (CPU-only system)
    \item \textbf{\hwpeer{}\textit{N+M}}: \bmac{} validator peer using N vCPUs and M \txvalidators{} in hardware (CPU+FPGA system)
\end{itemize}

\subsubsection{Varying Block Sizes}
Figure~\ref{blocksize_variation}a reports the energy efficiency of \swpeer{}8 and \hwpeer{}4+4 validator peers. We consider 8 vCPUs as the common multi-core configuration for \swpeer{} peer, while 4 vCPUs with 4 \txvalidators{} is the minimal \hwpeer{} peer configuration in our setup. Typically, 4 vCPUs are enough for \hwpeer{} peer because its software only commits blocks to disk-based ledger. It is evident that hardware accelerator significantly improves energy efficiency, up to 2.8$\times$ for block size 250 (15 vs. 42 tx/s/J). From a deeper analysis, we derive the following insights.

\textbf{Insight 1: }
The improvement in energy efficiency is quite notable for smaller block sizes (e.g., 10, 25 and 50). Small block sizes do not have enough transactions to fully utilize underlying compute resources and amortize the overhead of block processing. This is even more pertinent for hardware where internal pipelines do not fill up completely when block size is small. Therefore, one should choose a large enough block size keeping in mind the available compute resources for higher energy efficiency (e.g. 50 in this case).

\textbf{Insight 2: }
The energy efficiency saturates after a particular block size (e.g., after 50 and 100 for \swpeer{}8 and \hwpeer{}4+4 peers respectively). Once the block size is large enough to fully utilize the underlying compute resources and amortize the overhead of block processing, the validator peer achieves a steady state. Thus, increasing the block size further does not bring any improvements in throughput and/or energy consumption, however it does increase the memory footprint (storing larger blocks requires larger buffers in hardware). Therefore, one should choose a block size that just saturates energy efficiency of the validator peer.

\textbf{Insight 3: }
Figure~\ref{blocksize_variation}b shows the average power consumption of the two validator peers. The increase in power consumption across different block sizes is not significant which is expected because compute resources are unchanged. Interestingly, \hwpeer{}4+4 peer consumes 1.5$\times$ the power of \swpeer{}8 for block size 250. However, it also improves throughput by 2$\times$ (due to faster block validation, though not shown in the figure), which more than compensates for the increased power consumption resulting in higher energy efficiency. Therefore, looking at just the power consumption can be misleading and throughput/energy provides better insights.

\begin{figure}[t!]
\centering
\includegraphics[scale=0.5]{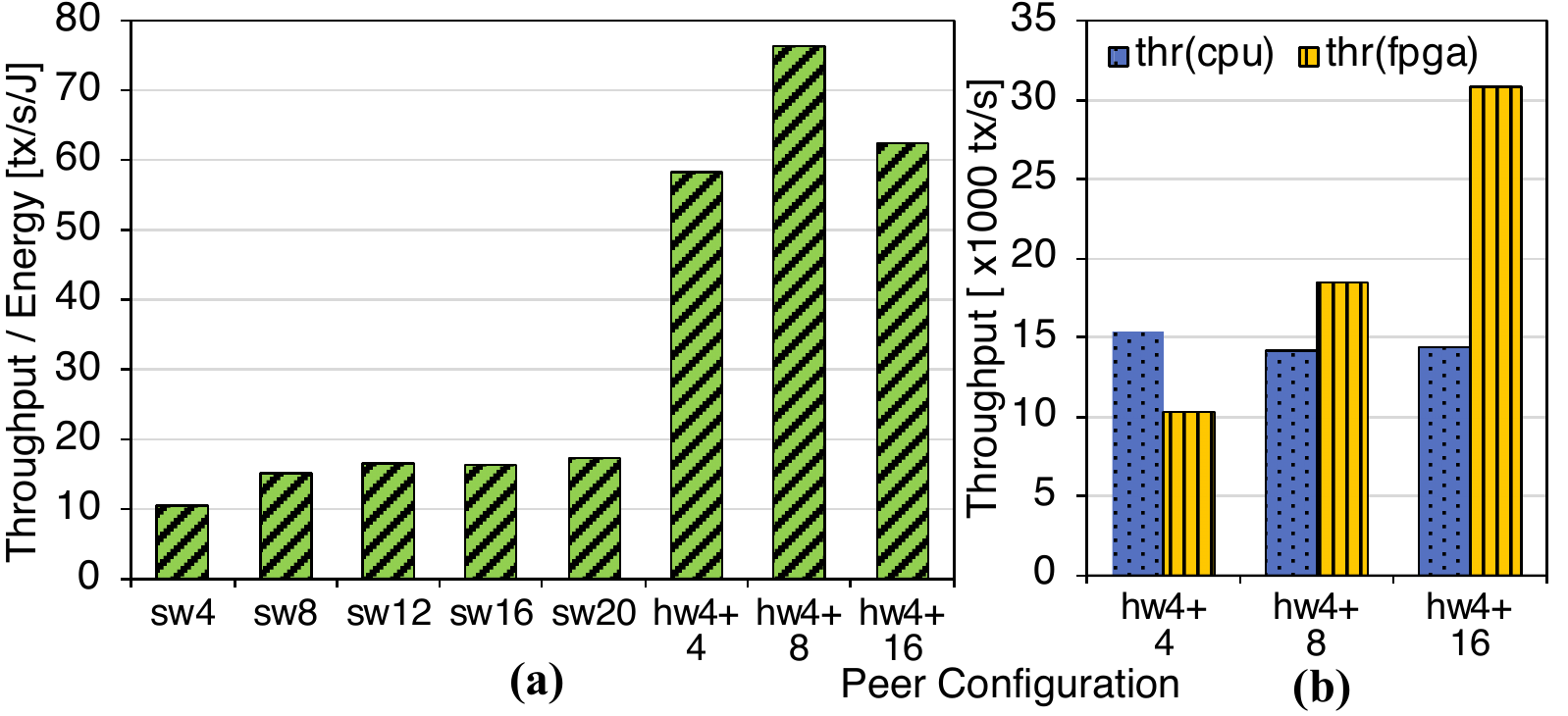}
\caption{(a) Energy efficiency, and (b) Throughput of various \swpeer{} and \hwpeer{} peer configurations.}
\label{cpu_variation}
\vspace{-3ex}
\end{figure}

\begin{figure}[t!]
\centering
\includegraphics[scale=0.5]{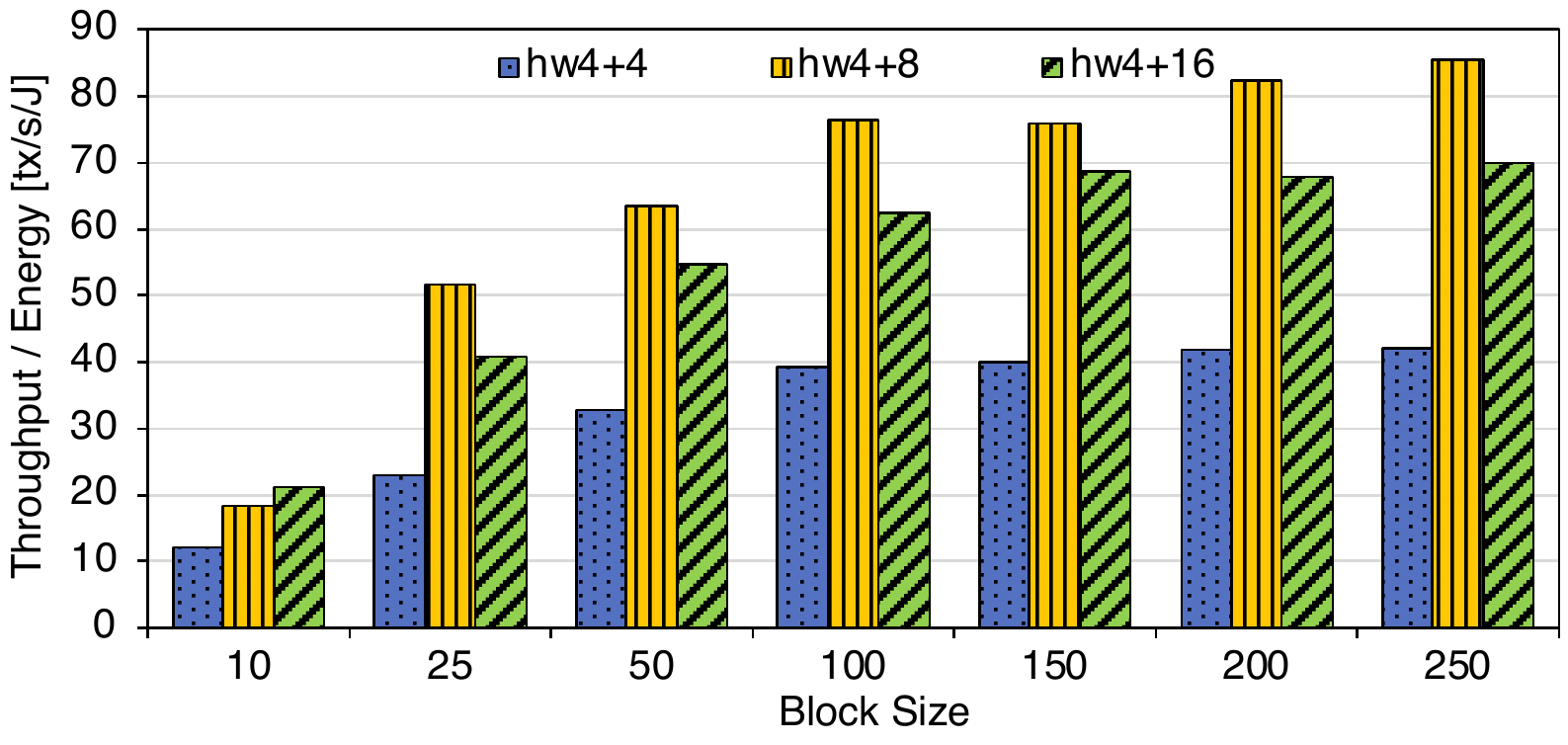}
\caption{Energy efficiency of various \hwpeer{} peer configurations.}
\label{hw_variation_thrperenergy}
\vspace{-4ex}
\end{figure}

\subsubsection{Varying Compute Resources}
Figure~\ref{cpu_variation}a shows the energy efficiency of the \swpeer{} and \hwpeer{} peers when the number of vCPUs and \txvalidators{} is changed with fixed block size of 100. In general, \hwpeer{} peer delivers much higher energy efficiency than \swpeer{} peer when more compute resources are added, from 58 tx/s/J to 76 tx/s/J vs. a maximum of 17 tx/s/J from \swpeer{}, resulting in 4.5$\times$ improvement. We deduce the following insights.

\textbf{Insight 1: }
The energy efficiency of \swpeer{} peer saturates after a particular number of vCPUs (e.g., 12). The validation phase only uses parallel threads for vscc operation, and mvcc and commit operations are executed sequentially without any pipelining, limiting the maximum throughput achievable. The extra power consumption from additional vCPUs is not compensated by notably higher throughput, and hence energy efficiency saturates. For example, \swpeer{}12, \swpeer{}16 and \swpeer{}20 peers consume 66W, 76W and 80W respectively, but deliver about the same 17 tx/s/J. Therefore, one should provision vCPUs that will just saturate energy efficiency of the validator peer but with minimal power consumption.

\textbf{Insight 2: }
The sudden drop in energy efficiency of \hwpeer{}4+16 vs. \hwpeer{}4+8 is quite unexpected. Since validation phase occurs in parallel on both the CPU and FPGA, we plot their individual throughputs in Figure~\ref{cpu_variation}b. The CPU throughput remains the same as the number of vCPUs does not change, while the FPGA throughput increases significantly from 4 to 16 \txvalidators{}. FPGA is the bottleneck in \hwpeer{}4+4 peer while CPU is the bottleneck in \hwpeer{}4+8 and \hwpeer{}4+16 peers. Thus, in \hwpeer{}4+16, power consumption from additional \txvalidators{} is not compensated at the validator peer level because of the much lower CPU throughput. Figure~\ref{hw_variation_thrperenergy} shows that the trend is same between \hwpeer{}4+8 and \hwpeer{}4+16 peers across all block sizes except 10, where CPU and FPGA throughputs are similar (not shown in figure). Therefore, one should match the compute power between the CPU and FPGA in order to achieve the highest energy efficiency.

\begin{figure}[t!]
\begin{subfigure}{.5\textwidth}
	\centering
	\includegraphics[scale=0.45]{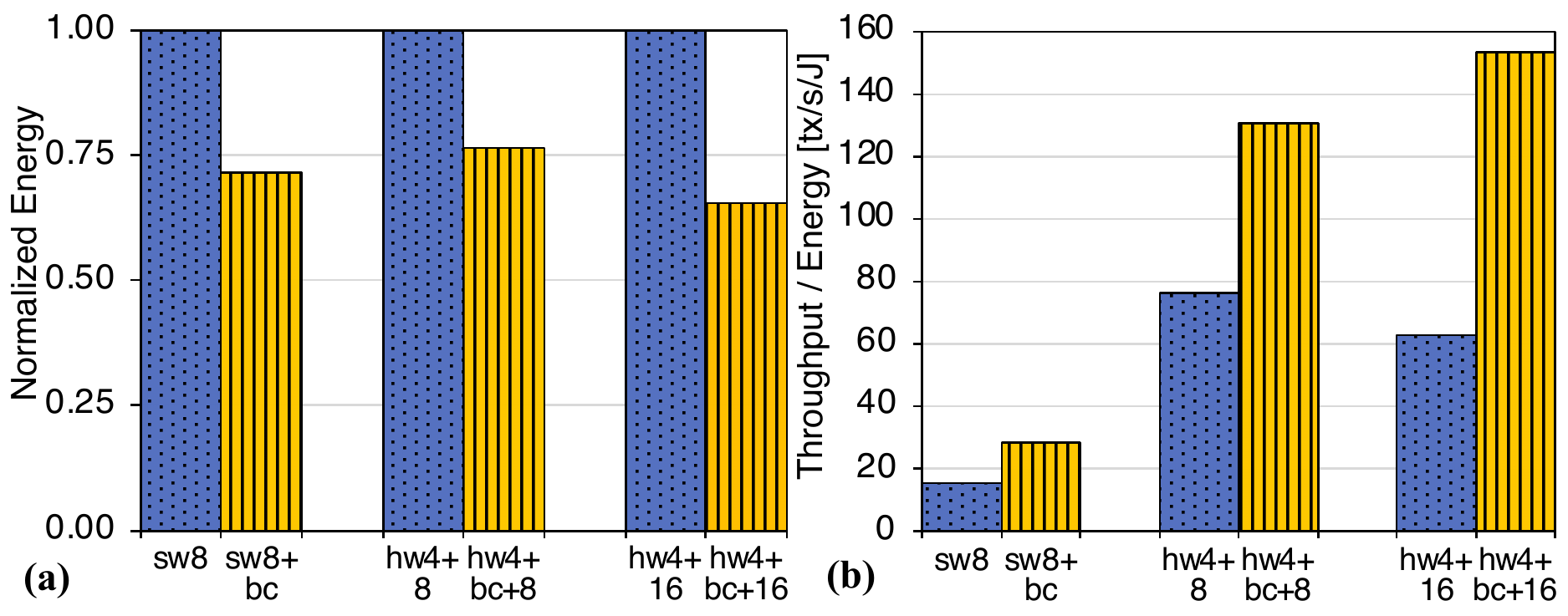}
\end{subfigure}

\begin{subfigure}{.5\textwidth}
	\centering
	\includegraphics[scale=0.45]{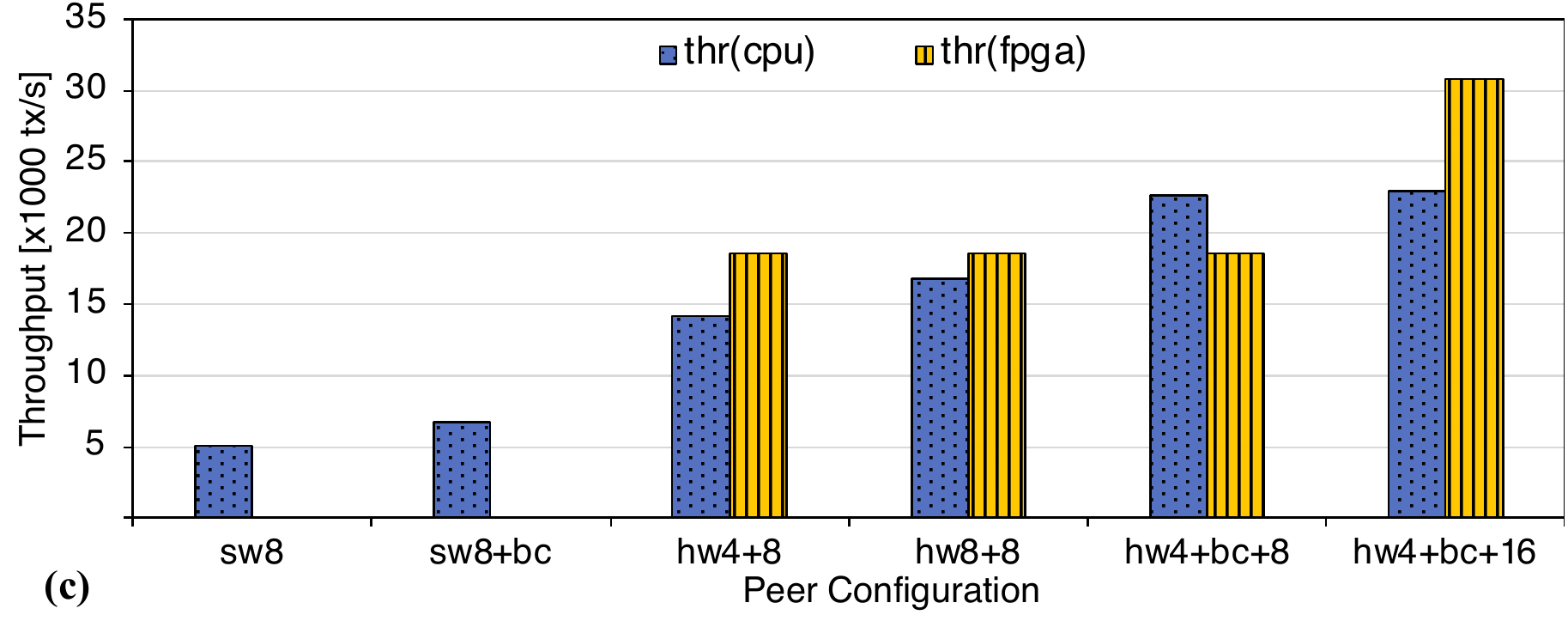}
\end{subfigure}

\caption{(a) Normalized energy consumption, (b) Energy efficiency, and (c) Throughput of various \swpeer{} and \hwpeer{} peer configurations.}
\label{config_variation}
\vspace{-4ex}
\end{figure}

\subsubsection{Software Block Cache}
We create two more variants of the validator peer to understand how software optimizations can affect its energy efficiency. We chose to implement the block cache from~\cite{Gorenflo2019FastFabric} in peer software because it is not yet part of Fabric v2.2 and has been shown to improve performance significantly~\cite{Gorenflo2019FastFabric,Thakkar2021ScalingBlockchains}.
\begin{itemize}
    \item \textbf{\swpeer{}\textit{N+bc}}: vanilla validator peer using N vCPUs and software block cache
    \item \textbf{\hwpeer{}\textit{N+bc+M}}: \bmac{} validator peer using N vCPUs and software block cache, and M \txvalidators{} in hardware
\end{itemize}

Figures~\ref{config_variation}a \&~\ref{config_variation}b report the normalized energy consumption and energy efficiency of various \swpeer{} and \hwpeer{} peers for a block size of 100. We do not include \hwpeer{}4+4 peer because FPGA is the bottleneck (Fig.~\ref{cpu_variation}b), thus software optimizations will not result in any noticeable improvement in its throughput or energy efficiency.

\textbf{Insight 1: }
The software block cache can bring in significant energy savings across both \swpeer{} and \hwpeer{} peers, up to 35\% in Fig.~\ref{config_variation}a. The peer software avoids many redundant operations because of the block cache, which results in faster block validation and hence lower energy consumption. The energy savings are more notable when CPU is the bottleneck in \hwpeer{} peer. Depending on the difference between CPU and FPGA throughput (Fig.~\ref{cpu_variation}b), FPGA can be idle for long periods because of the CPU. The use of block cache results in shorter FPGA idle periods, which means higher energy reduction due to FPGA's high power consumption. Figure~\ref{config_variation}b shows that the use of block cache results in the expected energy efficiency trend between \hwpeer{}4+8 and \hwpeer{}4+16 peers, in contrast to the unexpected trend in Fig.~\ref{cpu_variation}a. This is because CPU throughput is now closer to FPGA throughput which results in improved energy efficiency (compare \hwpeer{}4+16 in Figs.~\ref{cpu_variation}b \&~\ref{config_variation}c).

\textbf{Insight 2: }
Figure~\ref{config_variation}c reports how the CPU and FPGA throughputs are affected when resources/optimizations are changed. We show only the most interesting validator peer configurations here. Comparing \hwpeer{}4+bc+8 and \hwpeer{}8+8 with \hwpeer{}4+8 reveals that the use of block cache has much larger improvement in CPU throughput than adding more vCPUs (57\% vs. 18\%). Furthermore, CPU is still the bottleneck in \hwpeer{}8+8 peer. Therefore, one should carefully consider the interactions between hardware resources and software optimizations when provisioning a validator peer.

\subsubsection{Summary of Results}
We conclude that \hwpeer{}4+bc+16 peer synergistically benefits from its hardware resources and software optimizations, resulting in 153 tx/s/J which is the highest in all our experiments. This is a 10$\times$ improvement over 15 tx/s/J of \swpeer{}8 (representative of publicly available validator peer running on a common 8-core server). This means that \hwpeer{}4+bc+16 peer can deliver 10$\times$ more throughput than \swpeer{}8 while consuming the same amount of energy. In absolute terms, this translates to 23,000 tx/s  with a power consumption of 118W.

\vspace{-3ex}
\section{Related Work}\label{related_work}

Many recent works have proposed software and hardware optimizations to improve validator peer performance, such as parallel validation of transactions and/or pipelined execution of validation operations~\cite{Thakkar2018,Javaid2019,Gorenflo2019FastFabric,Kuhring2020,Thakkar2021ScalingBlockchains}, caching unmarshalled blocks~\cite{Gorenflo2019FastFabric} and offloading compute-intensive operations to specialized hardware~\cite{Javaid2022BlockchainMachine}. All these works only focus on performance improvements and overlook power/energy consumption of the validator peer, which is the focus of this paper. Many of these optimizations have already been incorporated into official Fabric v2.2 codebase (e.g., \cite{Thakkar2018,Javaid2019,Thakkar2021ScalingBlockchains}), so our evaluation already includes them. The software block cache~\cite{Gorenflo2019FastFabric} is not yet part of the official codebase, so we implemented it ourselves for evaluation. 

There is very little research on power/energy consumption of blockchains especially permissioned blockchains like Hyperledger Fabric. The work in~\cite{Sedlmeir2020BlockchainEnergy} estimated energy consumption of different blockchains, and emphasized that further detailed energy efficiency studies are needed especially for permissioned blockchains. The study in~\cite{Coroama2021BlockchainEnergy} proposed an analytical approach to estimate energy consumption of PoW-based blockchains and used Bitcoin as an example. Both these works are analytical in nature and do not use actual power/energy measurements as we have done in this paper. We are the first to conduct a comprehensive study of energy efficiency of Hyperledger Fabric's validator peer with actual power/energy measurements, and present insights for energy-aware provisioning of validator peers in a Fabric network.

\section{Conclusion}\label{conclusion}
In this paper, we proposed a power/energy measurement methodology for CPU and CPU+FPGA based systems in order to evaluate energy efficiency (throughput/energy) of Hyperledger Fabric's validator peer. We presented many useful insights from our comprehensive evaluation of a diverse set of validator peer configurations. We concluded that the right combination of hardware resources and software optimizations is essential for achieving highest energy efficiency. We achieved up to 153 tx/s/J compared to 15 tx/s/J of vanilla validator peer.

\section{Acknowledgments}
The authors thank Rajesh Panicker from NUS and Sundararajarao Mohan from AMD for their valuable support.

\bibliographystyle{IEEEtran}
\bibliography{icpads_paper.bbl}

\end{document}